\documentstyle[aps]{revtex}
\def\beq{\begin{equation}}
\def\eeq{\end{equation}}
\def\vecvar#1{\mbox{\boldmath$#1$}}
\newcounter{mycounta}
\newcounter{mycountb}
\def\beqa{\addtocounter{mycounta}{1}\begin{equation}}
\def\beqb{\addtocounter{mycountb}{1}\begin{equation}}
\def\kakko#1{\langle #1 \rangle}
\def\ac#1{{a^{(#1)}}^\dagger}
\def\aa#1{{a^{(#1)}}}

\def\admk{a^\dagger_{-\vecvar{k}}}
\def\amk{a_{-\vecvar{k}}}
\begin{document}
\draft
\preprint{TMUP-HEL-9906}
\title{Dynamical Pion Production via Parametric Resonance from 
Disoriented Chiral Condensate}
\author{Hideaki Hiro-Oka}
\address{Institute of Physics, Kitasato University\\
1-15-1 Kitasato Sagamihara, Kanagawa 228-8555, Japan}
\author{Hisakazu Minakata}
\address{Department of Physics, Tokyo Metropolitan University \\
1-1 Minami-Osawa, Hachioji, Tokyo 192-0397, Japan \\ and \\
Research Center for Cosmic Neutrinos,
Institute for Cosmic Ray Research, \\ 
University of Tokyo, Tanashi, Tokyo 188-8502, Japan}
\maketitle
\begin{abstract}
We discuss a dynamical mechanism of pion production from disoriented 
chiral condensates (DCC). It leads to an explosive production of pions 
via the parametric resonance mechanism, which is similar to the 
reheating mechanism in inflationary cosmology. 
Classically it is related with the instability in the solutions of 
the Mathieu equation and we explore the quantum aspects of the mechanism. 
We show that nonlinearities and back reactions can be ignorable for 
sufficiently long time under the small amplitude approximations of 
background $\sigma$ oscillations, which may be appropriate for the late
stage of nonequilibrium phase transition. It allows us to obtain 
an explicit quantum state of the produced pions and $\sigma$, the 
squeezed state of BCS type. Single particle distributions and two-pion 
correlation functions are computed within these approximations. 
The results obtained illuminate the characteristic features of 
multi-pion states produced through the parametric amplification 
mechanism. In particular, two-pion correlations of various charge 
combinations contain back-to-back correlations which cannot be masked 
by the identical particle interference effect.
We suggest that the parametric resonance mechanism might be a cause of 
the long lasting amplification of low momentum modes in linear sigma 
model simulations.

\end{abstract}

\section{Introduction}
We now believe in quantum chromodynamics (QCD) as {\it the} theory of 
strong interaction. Yet, the Centauro and anti-Centauro events that 
were found in cosmic ray experiments \cite{centauro} indicate some 
puzzling features. The events are characterized by large fluctuations 
in the ratio of number of neutral pions to that of charged pions. It 
appeared to be unlikely that such feature can be realized inside the 
conventional picture of hadronic interactions with "soft" QCD interaction, 
which implies more or less an independent emission of particles.

However, several authors formulated a scenario which explains such 
features of the events based on the idea of formation of chirally 
misaligned domains, the disoriented chiral condensate (DCC) 
\cite{preDCC,BJ}. It requires that a "hot" matter forms during 
hadronic collisions in which the chiral symmetry is restored, 
and then it rapidly cools down so that chiral orientation of 
the pion fields can align to a random direction different from 
that of the vacuum. If the low momentum mode of the pion fields are 
enhanced large domains of definite (but random in direction) 
isospin would form. Then, a large isospin, and therefore, charge 
fluctuations naturally results. 

The amplification of low momentum pion modes was demonstrated by 
Rajagopal and Wilczek \cite{RW93} by doing numerical simulation 
of the linear sigma model using the quench initial condition. 
A more explicit confirmation of the formation of large domains was 
provided by Asakawa, Huang and Wang \cite{AHW95} who computed pion-field 
correlation functions by using an elaborated simulation code which takes 
into account expansion along the longitudinal as well as the transverse 
direction. 

One should note, however, that the large charge-neutral fluctuation 
does not necessarily imply a random isotropic (in isospin space) 
rolling down from the top of the Mexican hat potential. As was shown 
by Greiner, Gong and M\"uller \cite{GGM93} any isosinglet multipion states 
approximately have the well known neutral fraction distribution $P(f)$ 
\begin{equation}
P(f)={1\over 2\sqrt{f}}.
\end{equation}
where 
\begin{equation}
f={N_{\pi^0}\over{N_{\pi^0}+N_{\pi^+}+N_{\pi^-}}}.
\end{equation}
We note that this distribution was in fact derived by only requiring  
the isospin singlet phase space \cite{RW92}. Therefore, 
the large charge-neutral fluctuation may merely imply isospin singlet 
DCC state, not a random isotropic rolling down.

The interpretation of the results of the simulation by Rajagopal and 
Wilczek \cite{RW93} also requires reexamination. They interpreted the 
origin of the amplification of the low momentum pion mode as due to 
instability of the Nambu-Goldstone modes of chirally symmetric field 
configurations at around the top of the Mexican hat. 
However, the enhancement of the 
low momentum pion modes actually take place with much longer time scale 
than that of rolling down motion, $\sim 1/m_{\sigma}$, as one can clearly 
see in Fig. 1 in their paper. Then, we need to identify certain mechanism 
which is responsible for the enhancement. The most likely candidate, to 
our opinion, is the parametric resonance mechanism \cite {LLmech,MM95}. 
The mechanism, possibly being tied up with a new instability that arises 
within the approximation scheme we will take, seems to be capable of 
explaining qualitative features of long lasting amplification of low 
momentum modes. (See sections 3 and 6.)
Some other aspects of formation mechanism of DCC were explored in Refs. 
\cite {Rand96,AMM,boyan1,Kayser}.

In a previous paper \cite{HM98}, we briefly explored the parametric 
resonance mechanism emphasizing its characteristic features in the 
two pion correlations, and its possible relevance as a hunting tool 
of DCC. 
In this paper we examine the mechanism in detail, hoping that it 
will shed some light to the origin of enhancement of the low momentum 
pion modes. 
We focus on the pion production at around the bottom of the 
Mexican hat potential. It may be an appropriate setting at least 
for the late stage of non-equilibrium phase transition which would 
result in the formation of coherent DCC domains. 

Furthermore, we concentrate on the case that the sigma model fields 
oscillate along the $\sigma$ direction. 
In fact, it is shown in the numerical simulation \cite{AHW95} that 
sigma model field configurations rapidly point to the $\sigma$ direction 
in a time scale of the order of $\sim 1/m_{\pi}$. 
Under the background oscillation of the $\sigma$ field the quantum 
fluctuations of the pion and the $\sigma$ fields can be excited.  
If the frequencies of background and quantum fluctuations match with 
each other the fluctuations are parametrically amplified, leading to 
an explosive pion production. 

We emphasize that the mechanism of pion production from DCC which we 
explore in this paper differs from that is usually adopted 
in the literature \cite{preDCC,BJ,RW93} as we mentioned earlier. 
In the conventional picture of pion emission from DCC, one assumes 
that the sigma model fields roll down along the direction 
(in most cases) different from the $\sigma$ direction. 
Then, the field configurations relax toward the $\sigma$ direction, 
the orientation of the QCD vacuum, which entails the coherent pion 
emission through the relaxation process. 
In contrast, pion production takes place even when the sigma model 
fields rolls down along the $\sigma$ direction in our parametric 
resonance mechanism. To indicate this point, we propose in the 
next section a set of approximations that leads to a simplified but 
a concrete model field theory that allows us to construct explicitly
the produced multi-pion quantum state.

We should remark that we do not claim that the parametric resonance 
is the whole story. Formation and decay of DCC is a complicated 
nonlinear phenomenon. The instability associated with the  
Nambu-Goldstone modes is indispensable ingredient for triggering the 
growth of the low momentum pion modes. But, we also stress that it is 
{\it not} enough either. In Sec. VI we will calculate the power spectrum 
of pion and sigma fields in our framework and attempt a qualitative 
comparison with the result obtained in Rajagopal-Wilczek's simulation.
It will give us a feeling on the question of to what extent the 
amplification of power is due to the parametric resonance mechanism.

Even though the parametric resonance mechanism is the cause of the 
long lasting amplification of low momentum modes we still lack 
understanding how the initial instability is mediated to it \cite {mina}. 
Leaving understanding of this last point to the future 
investigations our emphasis in this paper is that if the parametric 
resonance mechanism is operative at the late stage of the non-equilibrium 
phase transition it may give clear signatures which should be useful 
for its experimental hunting. 

In Sec. II, we introduce the linear sigma model, formulate the system 
with parametric resonance by introducing successive approximations in 
various stages, and make clear that to what extent the approximations 
are valid. 
In Sec. III, we quantize the system in a conventional way of quantizing 
a system of quantum fields, and clarify the relationship with the other 
formulation. 
In Sec. IV, we calculate the single particle distributions of pions and 
$\sigma$ to demonstrate a characteristic feature of the parametric 
resonance mechanism.
In Sec. V we discuss the two pion correlations which display another 
characteristic feature of the mechanism. We argue that they can be used 
as powerful experimental hunting tool of DCC. 
In Sec. VI we compute the power spectrum of pion and sigma fields 
and examine the qualitative features of the amplification of power 
in our framework.
In the last section we will summarize our investigation, 
give concluding remarks, and discuss the limitations of our framework.
In Appendix A, we summarize the feature of multiparticle states 
implied by the single mode squeezed state.
In Appendix B, we derive sum rule obeyed by isospin invariance and 
the iso-singlet nature of the multiparticle state within the framework
of single mode treatment.

\section{The Model and the Approximations}

The Lagrangian of the linear sigma model is given by 
\beq
{\cal L} = \frac{1}{2} \partial_{\mu}\phi_a \partial^{\mu}\phi_a
- \frac{\lambda}{4}(\phi_a\phi_a - v_0^2)^2
+ h\sigma.
\eeq
where $\phi_a =(\sigma, \vec{\pi})$. The typical values of the 
parameters which are relevant for phenomenology of QCD are: 
\beq
\lambda = 20, v_0 = 90\mbox{\rm MeV}, 
m_\pi=\sqrt{h\over v_0} = 140\mbox{MeV}.
\eeq
These values of the parameters will be used in the numerical 
analysis in later sections. 

As discussed in the previous section we deal with the pion production 
that takes place when the sigma model fields are at around the bottom 
of the Mexican hat potential, which may be realized in the time scale 
of $\sim 1/m_{\pi}$ after rolling down. 
We assume that at this stage the sigma model fields oscillate along 
the $\sigma$ direction. For clarity we construct a further 
simplified model by doing successive approximations in the linear 
sigma model that allow us to explicitly construct the produced 
multi-pion quantum state. 

We expand the sigma model fields at around the minimum of 
the Mexican hat potential along the $\sigma$ direction. 
\beq
\begin{array}{l}
\chi = \sigma - v,\\
\\
\vec \pi = \vec \pi.
\end{array}
\eeq
The linear sigma model Lagrangian, then, takes the form
\begin{equation}
\begin{array}{rl}
{\cal L}[\chi,\vec\pi] =& {\displaystyle\frac{1}{2}(\partial_{\mu} \chi)^2 
- {\lambda\over 2}(3v^2- v_0^2)\chi^2
- \lambda v \chi^3 - \frac{\lambda}{4}\chi^4
+ \frac{1}{2}(\partial_{\mu}\vec{\pi})^2}\\
&\\
&-\displaystyle \frac{1}{2}m_{\pi}^2\vec{\pi}^2
{\displaystyle- \lambda v \chi\vec{\pi}^2 
- {1\over 2}\lambda \chi^2\vec{\pi}^2
- \frac{1}{4}\lambda\vec{\pi}^4}.
\end{array}
\label{eqn:lagrangian}
\end{equation}

Following Mr\'owczy\'nski and M\"uller \cite{MM95}, 
we make a further approximation of decomposing $\sigma$ model fields 
into the classical time-dependent background fields and the quantum 
fluctuations around them 
\begin{equation}
\begin{array}{l}
\chi(\vecvar{x},t) 
= \chi_0(t) + \eta(\vecvar{x},t),\\
\\
\vec{\pi}(\vecvar{x},t) 
= \vec\pi_0(t) + \vec\xi(\vecvar{x},t).\label{decompo1}
\end{array}
\end{equation}
We have assumed the spatial homogeneity of the background fields. 
It may be a reasonable assumption if we work inside a single domain 
because the key feature of DCC is its coherence over a domain. Of course, 
it is a simplifying assumption in real physical situation, but is 
an inevitable one in order to make the treatment technically manageable.
Substituting (\ref{decompo1}) into Lagrangian (\ref{eqn:lagrangian}), 
we have
\beq
\begin{array}{rl}
&{\cal L}[\chi_0+\eta,\vec{\pi}_0+\vec\xi]={\cal L}[\chi_0,\vec{\pi}_0]\\&\\
&\kern 4mm\displaystyle +{1\over 2}(\partial\eta)^2-
{1\over 2}(m_\sigma^2+3\Sigma(t))\eta^2
-{1\over 2}\lambda\vec{\pi}_0^2\eta^2
-\lambda(\chi_0+v)\eta^3-{1\over 4}\lambda\eta^4\cr
&\cr
&\kern 4mm\displaystyle +{1\over 2}(\partial\vec\xi)^2
-{1\over 2}(m_\pi^2+\Sigma(t))\vec\xi^2
-{3\over 2}\lambda\vec{\pi}_0^2\vec\xi^2-\lambda\vec{\pi}_0\vec\xi^3-{1\over 4}\lambda\vec\xi^4\\
&\cr
&\kern 4mm\displaystyle -2\lambda v\vec{\pi}_0\eta\vec\xi-\lambda(\chi_0+v)\eta\vec\xi^2
-\lambda\vec{\pi}_0\vec\xi\eta^2-{1\over 2}\lambda\eta^2\vec\xi^2,
\label{new_L}
\end{array}
\eeq
where $m_{\sigma}=\sqrt{\lambda(3v^2-v_0^2)}$ and 
$v$ is
the minimum value of the $\sigma$ direction determined by the equation
$\lambda(\sigma^2-v_0^2)\sigma-h=0$. 
For small $h$, $v$ is given approximately by 
$v = v_0 + \frac{h}{2\lambda v^2}$. 
Due to the background field oscillation, a time-dependent term 
$\Sigma(t)$ arises in the mass terms of $\vec\xi$ and $\eta$ fluctuations 
and is given by $\Sigma=\lambda\chi_0(\chi_0+2v)$. 
The terms linearly proportional to fluctuations, of course, vanish due to 
the equations of motion of the background fields: 
\begin{equation}
\ddot{\chi}_0+\lambda\chi_0^3
+3\lambda v\chi_0^2+m_\sigma^2\chi_0+\lambda(\chi_0+v)\vec{\pi}_0^2 = 0,
\end{equation}
\begin{equation}
\ddot{\vec\pi}_0+m_\pi^2\vec{\pi}_0+2\lambda v\chi_0\vec{\pi}_0
+\lambda\chi_0^2\vec{\pi}_0+\lambda\vec{\pi}_0^3=0.
\end{equation}

We work with the Ansatz that the background fields 
oscillate along the $\sigma$ direction, and set $\vec{\pi}_0=0$ in 
the subsequent analyses. 
We further restrict ourselves into the regime, 
$|{\chi_0\over v}| \ll 1$, in which the nonlinear terms in 
$\chi_0$'s equation of motion are negligible. 
Then, the solution to the $\chi_0$'s equation of motion takes the 
harmonic form
\begin{equation}
\chi_0(t) =\tilde\chi_0 \cos(m_{\sigma}t + \varphi),\label{decompo2}
\end{equation}
where $\varphi$ is an initial phase which we set to vanish for simplicity.
In the numerical analysis performed in Sec.IV, we will use 
$|{\chi_0\over v}| = 0.05$ ($\times$ 1/2, 1, 2) for illustrative purpose.

We should note that the limitation of small amplitude oscillation
that we have imposed, $|{\chi_0\over v}| \ll 1$, restricts the 
validity of our treatment only to a qualitative level. 
Nevertheless, we believe that it is a meaningful starting point, and 
in fact it makes theoretical analysis transparent as we will see below.

We differ from Mr\'owczy\'nski and M\"uller \cite{MM95} by treating 
$\eta$ and $\vec\xi$ quantum mechanically. 
The equations of motion obeyed by fluctuations $\eta$ and $\vec\xi$ 
are given as
\begin{equation}
\begin{array}{l}
\partial^2{\eta}+m_\sigma^2\eta+3(2\lambda v \chi_0+\lambda\chi_0^2)\eta
+2\lambda(\chi_0+v)\eta^2+\lambda\eta^3=0,\\
\\
\partial^2{\vec\xi}+m_\pi^2\vec\xi+(2\lambda v \chi_0+\lambda\chi_0^2)\vec\xi
+\lambda\vec\xi^3=0.
\end{array}\label{eq_for_fluct}
\end{equation}
They are greatly simplified by the restriction of small oscillation 
amplitudes of background fields.  

Let us first focus on the pion fluctuations. The $\chi_0^2$ term is 
ignorable compared to $v\chi_0$ term because of the restriction of 
$|\frac{\chi_0}{v}| \ll 1$. The $v\chi_0$ term is also small 
compared with the mass term, but we shall keep it otherwise we lose the 
parametric resonance. The cubic term of $\vec\xi$ is negligible until 
the time scale 
\begin{equation}
z\sim {m_\sigma^2\over{4\lambda v \tilde{\chi}_0}}\ln\left(
\left(m_\sigma\over 2\right)^2-2\lambda v\tilde{\chi}_0\over{10\lambda}\right),
\end{equation}
where the dimensionless time $z$ is measured by $m_\sigma$ as 
$2z=m_\sigma t+\varphi+\pi$.
For instance, it can be numerically estimated as 
$z \sim 67$ for $|\frac{\chi_0}{v}| = 0.05$.

We have ignored the quantum back reactions to the pionic and sigma 
fluctuations due to particle production. 
It could be taken into account, e.g., by the Hartree-Fock approximation. 
Instead of going through this treatment, we estimate the time scale 
until which it can be negligible. It is given roughly as 
\begin{equation}
z\sim {m_\sigma^2\over{4\lambda v \tilde{\chi}_0}}\ln\left(
m_\sigma^2\tilde{\chi}_0\over {10\lambda(\tilde{\chi}_0+v)}\right),
\end{equation}
and the same numerical examination indicates that it is at $z \sim 32$. 

We emphasize, therefore, that under the approximation of small 
amplitude oscillation of background fields, ignoring nonlinear terms 
in the field fluctuations is a good approximation in a fairly long time 
even for such strong coupling as $\lambda = 20$. 

We end up with the equation of motion of sigma and pionic fluctuations: 
\begin{equation}
\left[ \frac{d^2}{dz^2} + A\sigma -2q_{\sigma}\cos(2z)\right]
\eta_{\vecvar k}^{(1)}(z) = 0,
\label{eqn:fluct1}
\end{equation}
\begin{equation}
\left[ \frac{d^2}{dz^2} + A_{\pi} -2q_{\pi}\cos(2z)\right]
\vec\xi_{\vecvar k}(z) = 0,
\label{eqn:fluct2}
\end{equation}
where
\begin{equation}
A_{\sigma} \equiv \frac{4({\vecvar k}^2+m_{\sigma}^2)}{m_{\sigma}^2}, 
\;\; q_{\pi} \equiv \frac{12\lambda v\tilde\chi_0}{m_\sigma^2},
\label{def_Aqsigma}
\end{equation}
and 
\begin{equation}
A_{\pi} \equiv \frac{4({\vecvar k}^2+m_{\pi}^2)}{m_{\sigma}^2}, 
\;\; q_{\pi} \equiv \frac{4\lambda v\tilde\chi_0}{m_\sigma^2},
\label{def_Aqpi}
\end{equation}
$\eta_{\vecvar{k}}$ and $\vec\xi_{{\vecvar k}}$ denote the Fourier 
components of 
the fields $\eta_{\vecvar{x}}$ and $\vec\xi(\vecvar{x})$, respectively.
The equations (\ref{eqn:fluct1}) and (\ref{eqn:fluct2}) are known as 
the Mathieu equation and are known to admit unstable solutions for a 
wide range of parameters of $A$'s and $q$'s. 
See, e.g., Ref. \cite{Mathieu}. Such unstable solution may be 
interpreted as explosive particle production under the background 
of oscillating $\chi_0$ fields. 

\section{Quantum Evolution of the System and the Squeezed State}

We discuss in this section the quantum evolution of the 
time-dependent states of $\vec\pi$ and $\sigma$ fluctuations.
Thanks to the restrictions and approximations introduced in the previous 
section, the structure of quantum states formed by $\eta$ and $\vec\xi$
quanta can be analyzed analytically. 

We start by writing the Hamiltonian in a form analogous to the 
harmonic oscillators but with time-dependent frequencies:
\beq
\begin{array}{rl}
H = \displaystyle\int d^3 k& \displaystyle\left[
\frac{1}{2} P_{\vecvar{k}} P_{-\vecvar{k}} + 
\frac{1}{2} \Omega_{\vecvar{k}}^{\sigma} (t)^2 
Q_{\vecvar{k}} Q_{-\vecvar{k}} \right]\\
&\\
&+ 
\displaystyle\sum_j \int d^3 k 
\left[
\frac{1}{2}P_{\vecvar{k}}^j P_{-\vecvar{k}}^j + 
\frac{1}{2}\Omega_{\vecvar{k}}^{\pi} (t)^2
Q_{\vecvar{k}}^j Q_{-\vecvar{k}}^j \right],
\label{effectiveH}
\end{array}
\eeq
where
\beq
\begin{array}{l}
\Omega_{\vecvar{k}}^{\sigma}(t)^2 
= {\vecvar k}^2 + m_{\sigma}^2 
+6\lambda v\chi_0(t),\\
\\
\Omega_{\vecvar{k}}^{\pi}(t)^2 
= {\vecvar k}^2 + m_{\pi}^2+2\lambda v\chi_0(t).
\end{array}\label{def_omega}
\eeq
The variable 
$Q_{\vecvar{k}}$ and $Q_{\vecvar k}^j$ in (\ref{effectiveH}) 
are defined by
\beq
\eta(\vecvar{x},t) = 
\displaystyle\int\frac{d^3 k}{(2\pi)^3}
e^{i\vecvar{k}\cdot \vecvar{x}} Q_{\vecvar{k}}(t),\quad
\xi^j(\vecvar{x},t) = 
\displaystyle\int\frac{d^3 k}{(2\pi)^3}
e^{i\vecvar{k}\cdot\vecvar{x}} Q_{\vecvar{k}}^j (t),
\label{defQ_k}
\eeq
and $P_{\vecvar k}$ and $P_{\vecvar k}^j$ are, as usual, 
the conjugate momenta of 
$Q_{\vecvar k}$ and $Q_{\vecvar k}^j$, respectively.
The index $j$ runs over 1 to 3, and we take the adjoint representation 
for the iso-triplet pion fields. For its concrete form, we refer 
Appendix B. One can see that the equations of motion (\ref{eqn:fluct1})
and (\ref{eqn:fluct2}) 
are derived from this Hamiltonian.

Within the small background oscillation we can take that 
$(\frac{\chi_0}{v}) < (\frac{m_\pi}{m_\sigma})^2$.
This restriction guarantees that the frequency 
$\Omega_{\vecvar{k}=0}^{\pi}$ is real, which is necessary in 
our present treatment which ignore quantum back reactions. 
If the background oscillation amplitude is larger than the critical 
value the system have an instability which we referred to as a new 
instability in Introduction. 
It definitely arises within our approximation and is indicative
of a remnant of the Rajagopal-Wilczek instability because it arises 
only for negative $\chi_0$
(which means toward central maximum of the Mexican hat).
Nevertheless, the new instability has a characteristic feature that is 
quite different from that of the Rajagopal-Wilczek's. 
The former can continue 
for long time while the latter lasts only a time scale of 
$1/m_\sigma$. Notice that both of the instabilities that we are 
talking about are the local instabilities, not the real global ones 
which totally destabilize the whole system. 

The nature of the instability, however, is not transparent to us.  
When one works with certain truncation in highly nonlinear systems 
one easily generate instabilities which may or may not be possessed 
by the original system. There are instabilities that one can easily 
understand their origin from simple consideration. For example, 
pion fields can have local instability by generating the winding 
motion encircling around the bottom of the wine bottle 
when the field energy of background $\sigma$ field oscillation with 
amplitude $\chi_0$ exceeds the symmetry breaking energy, that is, 
$\frac{1}{2} m_{\sigma}^2 \chi_0^2 > m_{\pi}^2 v^2$. It leads to the 
condition for onset of the instability
$\frac{\chi_0}{v} > \frac{m_{\pi}}{m_{\sigma}}$

But the instability we just encountered above occurs at 
much smaller amplitude, 
$(\frac{\chi_0}{v}) > (\frac{m_\pi}{m_\sigma})^2$.
To our knowledge there is no intuitive way of understanding the 
instability. It can be the artifact of our approximation and truncation 
scheme and, if this is the case, may be cured by taking nonlinearities 
and back reactions into account.

In the rest of the paper we will restrict ourselves into the small 
amplitude regime of the background $\sigma$ oscillation so that 
we are free from the new instability. 
In Sec. VII we will briefly discuss its role together with the 
parametric resonance as a possible candidate mechanism for understanding  
the long-lasting amplification of powers in sigma model simulations.

The quantum theory of harmonic oscillators with time-dependent 
frequencies were developed since sometime ago. 
In a previous paper \cite{HM98} we followed the formalism developed 
by Shtanov, Traschen and Brandenberger (STB)\cite{STB95}, but we shall 
present in this paper an equivalent but more conventional formalism of 
quantizing the same system. 
From the form of the Hamiltonian (\ref{effectiveH}), we can discuss
sigma and pion sector collectively; we treat below the sigma sector 
but with suppressing the superscript $\sigma$, and it can be regarded 
as that of pion sector in which the superscript $\pi$ as well as 
iso-spin index are suppressed.

We decompose the dynamical variables in momentum space in terms of
solutions of the equation of motions
\beq
\begin{array}{l}
Q_{\vecvar{k}}(t)=Q_1(t)a_{0\vecvar{k}}+Q_2(t)a^\dagger_{0-\vecvar{k}},\\
\\
P_{\vecvar{k}}(t)=\dot{Q}_1(t)a_{0-\vecvar{k}}
+\dot{Q}_2(t)a^\dagger_{0\vecvar{k}},
\label{def_Q}
\end{array}
\eeq
where $Q_1$ and $Q_2$ are the solutions to the equation of motion of $Q$
derived from the Hamiltonian,
\beq
\ddot{Q}_{\vecvar{k}}(t) + \Omega^2_{\vecvar{k}}(t)Q_{\vecvar{k}}(t) =0.
\label{eq_for_Q}
\eeq
Notice that $Q_2 = Q_1^*$.
$a_{0\vecvar{k}}$ and $a^\dagger_{0\vecvar{k}}$ are the annihilation and the 
creation operators, respectively. 
They are time-independent, i.e., they obey the Heisenberg equation of 
motion as
\beq
{da_{0\vecvar{k}}\over{dt}}\equiv{\partial a_{0\vecvar k}\over{\partial t}}+i[H,a_{0\vecvar k}]=0.
\eeq
The consistency condition between the commutation relations, i.e.,
\beq
\left[a_{0\vecvar{k}},~a^\dagger_{0\vecvar{k}}\right]=1,\qquad
\left[Q_{\vecvar{k}},~P_{\vecvar{k'}}\right]=i\delta(\vecvar{k-k'}).
\eeq
is achieved by the Wronskian condition
\beq
Q_1\dot{Q}_2-\dot{Q}_1Q_2=i.
\label{wronskian}
\eeq

Substituting (\ref{def_Q}) into (\ref{effectiveH}), we obtain the
following form
\begin{equation}
H_{\vecvar{k}}=\displaystyle\int d^3 k~~{1\over 2}\Biggl[
A_k a_{0\vecvar{k}}a_{0-\vecvar{k}} +
A_k^* a^\dagger_{0\vecvar{k}}a^\dagger_{0-\vecvar{k}}
+ B_k (a^\dagger_{0\vecvar{k}}a_{0\vecvar{k}} + a_{0\vecvar{k}}a^\dagger_{0\vecvar{k}})\,\Biggr],
\label{effH2}
\end{equation}
where the coefficients $A_{\vecvar k}$ and $B_{\vecvar k}$ are defined by
\begin{equation}
A_{\vecvar k} = \dot{Q}_1^2+\Omega^2_{\vecvar{k}}Q_1^2, \qquad
B_{\vecvar k} = \dot{Q}_1\dot{Q}_2+\Omega^2_{\vecvar{k}}Q_1Q_2.
\label{def_AB}
\end{equation}
Note that $A_{\vecvar k}$ is complex number, and that $Q_1$ and $Q_2$ are 
complex conjugate with each other. On the other hand, $B_{\vecvar k}$ is real 
as far as $\Omega_{\vecvar{k}}$ is real, which is the case owing to our 
restriction to small background oscillations. 

In order to diagonalize the Hamiltonian (\ref{effH2}), we 
introduce the Bogoliubov transformation as
\beq
\begin{array}{l}
a_{0\vecvar{k}}(t)=\alpha^*_{\vecvar k}(t)a_{\vecvar{k}}
-\beta^*_{\vecvar k}(t)a^\dagger_{-\vecvar{k}},\\
\\
a^\dagger_{0\vecvar{k}}(t)=\alpha_{\vecvar{k}}(t)
a^\dagger_{\vecvar{k}}-\beta_{\vecvar{k}}(t)a_{-\vecvar{k}},
\label{bogoliubov1}
\end{array}
\eeq
with
\beq
\alpha_{\vecvar{k}}(t)=e^{i\varphi_\alpha}\cosh\theta_{\vecvar{k}},\qquad
\beta_{\vecvar{k}}(t)=e^{i\varphi_\beta}\sinh\theta_{\vecvar{k}}.
\eeq
Choosing $\theta_k$ as 
\beq
\tanh 2\theta_{\vecvar k} ={|A_{\vecvar k}|\over B_{\vecvar k}},
\label{thetadiag}
\eeq
one can diagonalize the Hamiltonian into the form 
\beq
H_{\vecvar{k}}=\Omega_{\vecvar{k}}a^\dagger_{\vecvar{k}}(t)a_{\vecvar{k}}(t),
\label{effectiveH2}
\eeq 
where use has been made of the Wronskian relation (\ref{wronskian}) 
to show that the coefficient of (\ref{effectiveH2}) is in fact given 
by $\Omega_{\vecvar k}$.
Therefore, the operator $a_{\vecvar{k}}(t)$ defined in 
(\ref{bogoliubov1}) does represent annihilation operator 
of physical quanta at time $t$. 
Using (\ref{thetadiag}) with (\ref{def_AB}) is is easy to show that 
\beq
\cosh^2\theta_{\vecvar k} = \frac{B_{\vecvar k} + \Omega_{\vecvar k}}
{2\Omega_{\vecvar k}},\qquad 
\sinh^2\theta_{\vecvar k} = \frac{B_{\vecvar k} - \Omega_{\vecvar k}}
{2\Omega_{\vecvar k}}.
\label{chsh}
\eeq

In the above treatment, we have relied on the fact that amplitudes of 
the background oscillation is small so that $\Omega_{\vecvar{k}}$ is real.
If we remove this assumption the system experiences instability in 
a certain period of time. It is not totally unstable system due to 
the nature of the instability and it would be interesting to work out 
such system. However, we do not engage such business in the present paper. 
Instead, we argue that it is important to take account of back reactions 
in really addressing such problem. It would even be the case that the 
quantum back reaction entirely cures the instability. 

Let us move on to the analysis of the eigenstates for the effective 
Hamiltonian. We define the vacua $|0 \rangle$ and $|0(t) \rangle$ as 
\beq
a_{0\vecvar{k}}|0 \rangle = 0,\qquad
a_{\vecvar{k}}(t) |0(t) \rangle = 0,
\label{def_vacuum}
\eeq
and assume that the oscillator at $t=0$ is in the vacuum state 
$|0 \rangle$, i.e., $|0(t=0) \rangle = |0\rangle$. 
It leads to the initial conditions for $\alpha_{\vecvar k}(t)$ and 
$\beta_{\vecvar k}(t)$, 
\begin{equation}
|\alpha_{\vecvar{k}}(0)|=1, \qquad \beta_{\vecvar{k}}(0) =0.
\label{boundary}
\end{equation}
The physical quantum with momentum $\vecvar{k}$ is created (annihilated)
by $a_{\vecvar{k}}^{\dagger}(t)$ ($a_{\vecvar{k}}(t)$),
because the Hamiltonian is diagonalized by them at each time $t$.
As the system evolves, the state $|0\rangle$ is no longer the vacuum 
annihilated by $a_{\vecvar{k}}(t)$.

We rewrite the Bogoliubov transformation 
(\ref{bogoliubov1}) into the form of unitary transformation:
\beq
\begin{array}{l}
a_{0\vecvar{k}}=
e^{-i\varphi_\alpha}e^{-G_{\vecvar{k}}(t)}a_{\vecvar{k}}(t)
e^{G_{\vecvar{k}}(t)},\\
\\
a_{0\vecvar{k}}^\dagger=
e^{i\varphi_\alpha}e^{-G_{\vecvar{k}}(t)}a^\dagger_{\vecvar{k}}(t)
e^{G_{\vecvar{k}}(t)},
\end{array}
\label{bogoliubov2}
\eeq
where 
\beq
G_{\vecvar{k}}(t)=\theta_{\vecvar{k}}\left(a_{\vecvar{k}}^\dagger a_{-\vecvar{k}}^\dagger
-a_{\vecvar{k}}a_{-\vecvar{k}}\right).
\label{def_G}
\eeq
From the definitions of vacua (\ref{def_vacuum}) and of transformation 
(\ref{bogoliubov2}), we have
\beq
|0\rangle=e^{-G_{\vecvar k}(t)}|0(t)\rangle.
\eeq
After a short calculation\cite{AK95}, we obtain the vacuum state of the form
\begin{equation}
|0 \rangle = \prod_{\vecvar{k}} \displaystyle\frac{1}
{\sqrt{|\alpha_{\vecvar{k}}(t)|}}\mbox{exp} \left[
\frac{\beta^*_{\vecvar{k}}}{2\alpha^*_{\vecvar{k}}} a^{\dagger}_{\vecvar{k}}
(t) a^{\dagger}_{-\vecvar{k}}(t)\right] |0(t) \rangle.
\label{squeezed}
\end{equation}
The state is known as the squeezed state and widely used in quantum 
optics \cite{optics}. 
A significant feature of (\ref{squeezed}) is that it has pairing 
correlation between back-to-back, namely $\vecvar{k}$ and $-\vecvar{k}$, 
modes, and 
we will refer to it as the squeezed state of BCS type in this paper.  
In the next two sections, we will discuss, in detail, what are the 
characteristic features of particle production due to the parametric 
resonance mechanism. 
Earlier discussions of the squeezed state in the context of DCC may be 
found in \cite{AK95,DH96}.
We also note that the particle production via the parametric resonance 
has been extensively discussed in the context of reheating in the 
inflationary universe \cite{linde}-\nocite{boyan2}\cite{yoshimura}, 
and a mechanism for producing superheavy dark matter \cite{dark}.

Before closing this section, we would like to comment on the 
relationship between our formalism and the one by Shtanov, 
Traschen, and Brandenberger \cite{STB95} which can provide 
description of quantum systems with parametric resonances. 
Since we have adopted their formalism in our previous paper, 
we will try to compare and clarify the relation between various 
quantities that appear in both formalism. 

In short, there is a simple relationship between two formalisms: 
they start from the time-dependent operator (our $a_{\vecvar k}(t)$) which 
diagonalizes the Hamiltonian and then Bogoliubov transform it to the 
time-independent operator $a_{0\vecvar k}$. 
Therefore, such a correspondence is likely to exists. 

In order to make the correspondence more precise, we write down 
the equations obeyed by the coefficients of the Bogoliubov 
transformation in both formalisms. 
In STB formalism, they read 
\begin{equation}
\begin{array}{rl}
\dot{\alpha}_{\vecvar k}(t)=&\displaystyle {\dot{\Omega}_{\vecvar k}
\over{2\Omega_{\vecvar k}}}\beta_{\vecvar k}
\cdot e^{2i\int^t\Omega_{\vecvar k}dt'},\\
&\\
\dot{\beta}_{\vecvar k}(t)=&\displaystyle {\dot{\Omega}_{\vecvar k}
\over{2\Omega_{\vecvar k}}}\alpha_{\vecvar k}
\cdot e^{-2i\int^t\Omega_{\vecvar k}dt'}.\\
\end{array}\label{eq_for_coefficient}
\end{equation}
On the other hand,
the equations of $\alpha_{\vecvar k}$ and $\beta_{\vecvar k}$ 
in our formalism are given by 
\begin{equation}
\begin{array}{rl}
\dot{\alpha}_{\vecvar k}(t)&
\displaystyle ={\dot{\Omega}_{\vecvar k}\over {2\Omega_{\vecvar k}}}
\beta_{\vecvar k}\cdot
{e^{2i\varphi_\alpha}\over{\dot{\Omega}_{\vecvar k} A_{\vecvar k}}}
\biggl[2i\dot{\varphi}_\alpha(B_{\vecvar k}+\Omega_{\vecvar k})
\Omega_{\vecvar k}+\dot{B_{\vecvar k}}\Omega_{\vecvar k}
-B_{\vecvar k}\dot{\Omega}_{\vecvar k}\biggr],\\
&\\
\dot{\beta}_{\vecvar k}(t)&
\displaystyle ={\dot{\Omega}_{\vecvar k}\over {2\Omega_{\vecvar k}}}
\alpha_{\vecvar k}\cdot
{e^{-2i\varphi_\alpha}\over{\dot{\Omega}_{\vecvar k}A_{\vecvar k}^*}}
\biggl[2i\dot{\varphi}_\beta(B_{\vecvar k}-\Omega_{\vecvar k})
\Omega_{\vecvar k}+\dot{B_{\vecvar k}}\Omega_{\vecvar k} 
-B_{\vecvar k}\dot{\Omega}_{\vecvar k}\biggr].\\
\end{array}\label{eq_for_coefficient2}
\end{equation}
Above two sets of equations are consistent with each other provided 
that the phases of the coefficients satisfy the transformation rule:
\begin{equation}
\begin{array}{l}
\displaystyle e^{i\varphi_\alpha}=e^{i\int^tdt'\Omega_{\vecvar{k}}(t')}
\left({i\dot{Q}_1+\Omega_{\vecvar{k}}Q_1\over
{-i\dot{Q}_2+\Omega_{\vecvar{k}}Q_2}}\right)^{1/2},\\
\displaystyle e^{i\varphi_\beta}=e^{-i\int^tdt'\Omega_{\vecvar{k}}(t')}
\left({-i\dot{Q}_1+\Omega_{\vecvar{k}}Q_1\over
{i\dot{Q}_2+\Omega_{\vecvar{k}}Q_2}}\right)^{1/2}.
\label{consistency}
\end{array}
\end{equation}
Namely, the phase parts, which are undetermined in our present formalism,
are chosen in a particular way in the STB formalism.
Notice that the right-hand-sides of (\ref{consistency}) are pure phases 
and hence they are consistent.

\section{Single Particle Distributions}

Let us discuss the qualitative features of the single particle 
distributions of particles produced to explore the parametric 
resonance mechanism.
The multi pion and sigma state in our mechanism is given by 
\begin{equation}
|0 \rangle^{\sigma} \otimes |0 \rangle^{{\pi}} 
\equiv |\psi \rangle,
\label{ketpsi}
\end{equation}
where $|0 \rangle$ stands for the BCS type squeezed state 
(\ref{squeezed}).
The factorized form in (\ref{ketpsi}) stems from the linear 
approximations described in Sec. II.

The average number of quanta 
$\langle n \rangle_{\vecvar{k}}(t)$
with momentum $\vecvar{k}$ 
at time $t$ is given by
\beq
\langle n \rangle_{\vecvar{k}}(t) = 
\langle 0|a_{\vecvar{k}}^{\dagger}(t)a_{\vecvar{k}}(t)|0\rangle.
\label{one-point}
\eeq
We suppress in this section the isospin index, since the single particle 
distributions is independent of isospin.
Given the squeezed state (\ref{squeezed}), it can be readily 
shown to be 
\beq
\langle n \rangle_{\vecvar{k}}(t) = 
|\beta_{\vecvar{k}}|^2 = 
\frac{B_{\vecvar k} - \Omega_{\vecvar k}}{2\Omega_{\vecvar k}},
\label{lbeta}
\eeq
by using (\ref{chsh}) with $B_{\vecvar k}$ defined in (\ref{def_AB}).

Before presenting results of numerical computation, we discuss 
what would be the characteristic feature of the single particle 
distributions. It is natural to expect that an enhancement occurs 
due to the parametric resonance mechanism.
With physical values of the parameters, we take the $q$ parameters 
defined in (\ref{def_Aqsigma}) and (\ref{def_Aqpi}) are given by
\beq
q_\pi = 2.0 \left(\frac{\chi_0}{v}\right), \qquad
q_\sigma = 6.0 \left(\frac{\chi_0}{v}\right).
\eeq
Numerically, $q_\pi$ and $q_\sigma$ are less than 
$\sim 0.1$ and $\sim 0.3$, respectively, 
under the present approximation 
$(\frac{\chi_0}{v}) < (\frac{m_\pi}{m_\sigma})^2$.
Therefore, we are working with narrow resonance approximation. 
One should keep in mind, however, that the narrow resonance 
approximation may not hold in the real world. 

At such small $q$ parameter, the resonance occurs in the narrow 
bands at around the discrete values of $A$, $A=n^2, (n=1, 2, 3, ...)$.
The first resonance takes place for pion at $A_\pi=1$ and it 
correspond to $k=276$ MeV. For sigma, the first resonance 
takes place at $A_\sigma=4$ at zero momentum. The second resonance 
would be located at $A_\pi=4$ and at $A_\sigma=9$ which imply 
$k=603$ MeV and $k=692$ MeV for pion and sigma, respectively. 
Since the parameter $A$ is independent of $(\frac{\chi_0}{v})$, we 
expect that the peak position is stable against varying $\chi_0$, 
as will be demonstrated below.

We solve the Mathieu equation (\ref{eq_for_Q}) subject to the 
boundary condition (\ref{boundary}), and then compute 
$\langle n \rangle_{\vecvar{k}}(t)$.
In Fig. 1-3 
we plot the single particle momentum distributions of 
pion as a function of momentum $\vecvar k$ and dimensionless time $z$ with 
background 
oscillation parameters $\chi_0\over v$ = 0.025, 0.05, and 0.1. 
We observe that a prominent peak exists at the right position for 
the first resonance in every three figures indicating 
the parametric resonance enhancement. 
In Fig. 3 the momentum range 
$|\vecvar{k}| < \sqrt{0.1m_{\sigma}^2 - m_{\pi}^2}$ 
in which an exponential instability exists (due to imaginary frequency) 
is cut off.

In Fig.\ref{s-amp0.05}, we give the single particle momentum distributions of 
sigma at $\chi_0\over v$ = 0.05. The first resonance, that is expected 
at $\vecvar k=0$, is barely seen in the figure. The second resonance is 
invisible both in pion and sigma distributions.

\section{Two Pion Correlations}

We now discuss two particle distributions and correlations between 
pions. The quantum pion and sigma state is summarized 
in (\ref{ketpsi}) under the present approximations.
We focus on the pion distributions of various charge states.
Because of the factorized form of (\ref{squeezed}) and (\ref{ketpsi}) 
$n$-particle distributions of $\sigma$ can readily be obtained from 
that of $\pi^{0}$ by an appropriate change of the parameters.
It is true that the interpretation of our results on sigma particles 
in real experiments is not very transparent. Furthermore, when 
the $\sigma\rightarrow\pi\pi$ coupling is turned on the $\sigma$ 
particles do affect the two pion correlations. Discussion of 
this point is, however, beyond the scope of this work. 

Two pion distributions are defined as
\beq
\langle\pi_{\vecvar{k}}^i\pi_{\vecvar{k'}}^j\rangle\equiv
\langle \psi| a^{i\dagger}_{\vecvar{k}}(t) a^i_{\vecvar{k}}(t) 
a^{j\dagger}_{\vecvar{k'}}(t) a^j_{\vecvar{k'}}(t) | \psi\rangle,
\eeq
Because of the factorized form of (\ref{ketpsi}), 
there are no correlations between pion and sigma. Also, because of the 
factorized form of the squeezed state (\ref{squeezed}) in momentum 
space, the nontrivial two particle correlations exist only between 
modes of identical momentum ($\vecvar k$ and $\vecvar k$), or between 
back-to-back momentum ($\vecvar k$ and $-\vecvar k$) configurations.
One of the most important feature of the state (\ref{ketpsi}) is 
that it is the iso-singlet state. It comes from the fact that the 
frequency $\Omega_{\vecvar k}^{\pi}$ is isospin singlet. As we 
discuss below, it will give us nontrivial constraints on two 
pion correlations of various charge combinations. 
At the same time it also generates the large charge-neutral fluctuations 
as discussed in Sec. I. 

In the computation of these quantities, we further recognize that 
the two-pion distributions at zero-momentum can {\it not} be obtained 
by taking the smooth limit $\vecvar{k}\rightarrow 0$ of the 
expressions of either identical or back-to-back momentum
configurations. It is because $a_{\vecvar{k}=0}$ and 
$a^{\dagger}_{\vecvar{k}=0}$ 
do not commute, whereas $a_{\vecvar{k}}$ and $a^{\dagger}_{-\vecvar{k}}$ 
do commute for $\vecvar{k} \neq 0$. Therefore, we have to compute 
three types of the two-pion distributions separately. 

We briefly describe some details of the computation of two pion 
distributions. It is convenient to use adjoint representation 
operator $a^i_{\vecvar k}$ for this purpose. 
The two pion distribution of the same momentum $\vecvar k$
goes as follows:
\beq
\begin{array}{rl}
\langle\pi_{\vecvar{k}}\pi_{\vecvar{k}}\rangle&=
\kakko{a^\dagger_{\vecvar{k}} a_{\vecvar{k}}
a^\dagger_{\vecvar{k}} a_{\vecvar{k}}}\\
&\\
&=\kakko{a^\dagger_{\vecvar{k}}a^\dagger_{\vecvar{k}}
a_{\vecvar{k}}a_{\vecvar{k}}}+\kakko{a^\dagger_{\vecvar{k}}a_{\vecvar{k}}}\\
&\\
&=\left|{\beta_{\vecvar{k}}\over\alpha_{\vecvar{k}}}\right|^4
\kakko{a_{-\vecvar{k}}a_{-\vecvar{k}}
a^\dagger_{-\vecvar{k}}a^\dagger_{-\vecvar{k}}}+
\left|\beta_{\vecvar{k}}\right|^2\\
&\\
&=\left|{\beta_{\vecvar{k}}\over\alpha_{\vecvar{k}}}\right|^4\left(
\kakko{\admk\amk\amk\admk}+2\kakko{\amk\admk}\right)+
\left|\beta_{\vecvar{k}}\right|^2\\
&\\
&=\left|{\beta_{\vecvar{k}}\over\alpha_{\vecvar{k}}}\right|^4\left(
\kakko{\admk\amk\admk\amk}+3\kakko{\admk\amk}+2\right)+
\left|\beta_{\vecvar{k}}\right|^2\\
&\\
&=\left|{\beta_{\vecvar{k}}\over\alpha_{\vecvar{k}}}\right|^4\left(
\kakko{\pi_{-\vecvar{k}}\pi_{-\vecvar{k}}}+3\left|\beta_{\vecvar{k}}\right|^2
+2\right)+\left|\beta_{\vecvar{k}}\right|^2,
\end{array}
\label{twopi+}
\eeq
where the third equality stems from the relation
\beq
a_{\vecvar k}| \psi\rangle = 
\frac{\beta^*_{\vecvar k}}{\alpha^*_{\vecvar k}}\cdot 
a^\dagger_{- \vecvar k}| \psi\rangle.
\eeq
Similarly, $\kakko{\pi_{-{k}}\pi_{-{k}}}$ is given as
\beq
\kakko{\pi_{-\vecvar{k}}\pi_{-\vecvar{k}}}=
\left|{\beta_{\vecvar{k}}\over\alpha_{\vecvar{k}}}\right|^4\left(
\kakko{\pi_{\vecvar{k}}\pi_{\vecvar{k}}}+3\left|\beta_{\vecvar{k}}\right|^2
+2\right)+\left|\beta_{\vecvar{k}}\right|^2.
\label{twopi-}
\eeq
By solving the coupled equations (\ref{twopi+}) and (\ref{twopi-}),
we obtain two pion distributions of $j$-th isospin component as 
\beq
\kakko{\pi^j_{\vecvar{k}}\pi^j_{\vecvar{k}}}=
\kakko{a^\dagger_{\vecvar{k}} a_{\vecvar{k}}a^\dagger_{\vecvar{k}} 
a_{\vecvar{k}}}=
\left|\beta_{\vecvar{k}}\right|^2\left(2\left|\beta_{\vecvar{k}}\right|^2
+1\right)=
\kakko{\pi^j_{{- \vecvar k}}\pi^j_{{-\vecvar k}}}.
\eeq
Via analogous ways, one can compute the following expressions 
of the elements of nontrivial two pion correlations. 

For the same $\vecvar{k}\ne 0$:
\beq
\begin{array}{l}
\kakko{a_{i\vecvar{k}}^\dagger a_{i\vecvar{k}}}=
\left|\beta_{\vecvar{k}}\right|^2,\\
\\
\kakko{a_{i\vecvar{k}}^\dagger a_{i\vecvar{k}}
a_{j\vecvar{k}}^\dagger a_{j\vecvar{k}}}=\left|\beta_{\vecvar{k}}\right|^4 
~(\mbox{\rm for } i \ne j),\\
\\
\kakko{a_{i\vecvar{k}}^\dagger a_{i\vecvar{k}}
a_{i\vecvar{k}}^\dagger a_{i\vecvar{k}}}=\left|\beta_{\vecvar{k}}\right|^2
\left( 2\left|\beta_{\vecvar{k}}\right|^2+1\right),\\
\\
\kakko{a_{i\vecvar{k}}^\dagger a_{i\vecvar{k}}^\dagger
a_{j\vecvar{k}} a_{j\vecvar{k}}}=0 ~~(\mbox{for }i \ne j).
\end{array}
\eeq

For opposite $\vecvar{k}\ne 0$:
\beq
\begin{array}{l}
\kakko{a_{i\vecvar{k}}^\dagger a_{i-\vecvar{k}}}=0,\\
\\
\kakko{a_{i\vecvar{k}}^\dagger a_{i\vecvar{k}}
a_{j-\vecvar{k}}^\dagger a_{j-\vecvar{k}}}=\left|\beta_{\vecvar{k}}\right|^4 
~(\mbox{\rm for } i \ne j),\\
\\
\kakko{a_{i\vecvar{k}}^\dagger a_{i\vecvar{k}}
a_{i-\vecvar{k}}^\dagger a_{i-\vecvar{k}}}=\left|\beta_{\vecvar{k}}\right|^2
\left( 2\left|\beta_{\vecvar{k}}\right|^2+1\right),\\
\\
\kakko{a_{i\vecvar{k}}^\dagger a_{j\vecvar{k}}
a_{i-\vecvar{k}}^\dagger a_{j-\vecvar{k}}}=\left|\beta_{\vecvar{k}}\right|^2
\left( \left|\beta_{\vecvar{k}}\right|^2+1\right).
\end{array}
\eeq

By using these formulae, it is straightforward to compute the two pion 
distributions of definite isospin states. We transform them into the 
ones written by pion charge states, which are more suitable to gain 
insights for the experiments. We will describe in Appendix B 
the relationship between the adjoint representation pion operator 
(with which we have worked) and the charge-state operator, the 
(slightly modified) Horn-Silver representation
\beq
\begin{array}{ll}
a^{(+)}_{\vecvar{k}}= -\displaystyle{1\over\sqrt{2}}\left(a^1_{\vecvar{k}}
- ia^2_{\vecvar{k}}\right),\\&\\
a^{(-)}_{\vecvar{k}}= \displaystyle{1\over\sqrt{2}}\left(a^1_{\vecvar{k}}
+ ia^2_{\vecvar{k}}\right),\\&\\
a^{(0)}_{\vecvar{k}}= a^3_{\vecvar{k}},
\end{array}
\label{chargestate}
\eeq
where $a^{(\pm)}_{\vecvar{k}}$ and $a^{(0)}_{\vecvar{k}}$ stand for the 
annihilation operators of $\pi^\pm$ and $\pi^0$ with momentum $\vecvar k$, 
respectively. 
We express the two pion distributions of various charge combinations 
in the form of the ratio $R$ of them to the single pion distributions as 
defined by

\beq
R_{\vecvar{k}_1, \vecvar {k}_2}(p, q) = 
\frac{\langle\pi^{(p)}_{{\vecvar k_1}}\pi^{(q)}_{{\vecvar k_2}}\rangle}
{\langle\pi^{(p)}_{{\vecvar k_1}}\rangle \langle\pi^{(q)}_{{\vecvar k_2}}\rangle }, 
\eeq
where $p$ and $q$ stand for $\pm$ and 0. 
They read:

\noindent
(a) identical momentum distribution:
\beq
\begin{array}{l}
R_{\vecvar{k}, \vecvar {k}}(+, +) =
R_{\vecvar{k}, \vecvar {k}}(0, 0) =
2 + \displaystyle\frac{1}{\langle n \rangle_{\vecvar{k}}},\\
\\
R_{\vecvar{k}, \vecvar {k}}(+, -) = 1. 
\end{array}
\eeq

\noindent
(b) back-to-back momentum distribution: 
\beq
\begin{array}{l}
R_{\vecvar{k}, -\vecvar {k}}(+, -) =
R_{\vecvar{k}, -\vecvar {k}}(0, 0) =
2 + \displaystyle\frac{1}{\langle n \rangle_{\vecvar{k}}},\\
\\
R_{\vecvar{k}, -\vecvar {k}}(+, +) = 1.
\end{array}
\eeq

\noindent
(c) zero-momentum distribution: 
\beq
\begin{array}{l}
R_{\vecvar{k}=0, \vecvar {k}=0}(+, +) =
R_{\vecvar{k}=0, \vecvar {k}=0}(+, -) =
2 + \displaystyle\frac{1}{\langle n \rangle_{\vecvar{k}=0}},\\
\\
R_{\vecvar{k}=0, \vecvar {k}=0}(0, 0) =
3 + \displaystyle\frac{2}{\langle n \rangle_{\vecvar{k}=0}}.
\end{array}
\eeq

As an another measure for two pion correlations, we define 
the correlation function in the following way:
\begin{equation}
C(\pi^a_{\vecvar{k}}, \pi^b_{\vecvar{k'}}) \equiv 
\langle \pi^a_{\vecvar{k}}, \pi^b_{\vecvar{k'}} \rangle -
\delta_{ab}\delta_{\vecvar{k},\vecvar{k'}} \langle\pi^a_{\vecvar{k}} \rangle - 
\langle \pi^a_{\vecvar{k}} \rangle \langle \pi^b_{\vecvar{k'}}\rangle.
\end{equation}
Then, it is straightforward to obtain the following results. 

\noindent
(a) identical momentum correlations: 
\beq
\begin{array}{l}
C(\pi^{+}_{\vecvar{k}},\pi^{+}_{\vecvar{k}}) =
C(\pi^0_{\vecvar{k}}, \pi^0_{\vecvar{k}}) = 
\langle n \rangle^2_{\vecvar{k}},\\
\\
C(\pi^{+}_{\vecvar{k}}, \pi^{-}_{\vecvar{k}}) = 0.
\end{array}
\eeq
\noindent
(b) back-to-back momentum correlations:
\beq
\begin{array}{l}
C(\pi^{+}_{\vecvar{k}}, \pi^{-}_{-\vecvar{k}}) =
C(\pi^0_{\vecvar{k}}, \pi^0_{-\vecvar{k}}) = 
\langle n_{\vecvar{k}} \rangle
\left(\langle n \rangle_{\vecvar{k}} +1 \right),\\
\\
C(\pi^{+}_{\vecvar{k}}, \pi^{+}_{-\vecvar{k}}) = 0.
\end{array}
\eeq

\noindent
(c) zero-momentum correlations: 
\beq
\begin{array}{l}
C(\pi^{+}_{\vecvar{k}=0}, \pi^{+}_{\vecvar{k}=0}) =
\langle n \rangle^2_{\vecvar{k}=0},\\
\\
C(\pi^{+}_{\vecvar{k}=0}, \pi^{-}_{\vecvar{k}=0}) = 
\langle n \rangle_{\vecvar{k}=0}
\left(\langle n \rangle_{\vecvar{k}=0} +1 \right), \\
\\
C(\pi^{0}_{\vecvar{k}=0}, \pi^{0}_{\vecvar{k}=0}) = 
\langle n \rangle_{\vecvar{k}=0}
\left(2\langle n \rangle_{\vecvar{k}=0} +1 \right). 
\end{array}
\eeq
As will be discussed in Appendix B, the features of the 
correlation functions in (c) can be understood partly as a 
consequence of the isospin singlet nature of (\ref {ketpsi}).

The identical and back-to-back momentum correlations are depicted in figures
\ref{csame} and \ref{copp}.
In these figures, the amplitude parameter is taken as $\chi_0/v=0.05$ and
it is integrated over dimensionless time $z$ from 0 to 10, 
which corresponds to the period of time from $t=0$ to 6.5 fm.

\section{Power Spectrum and Possible Mechanisms for Amplification of 
Low-Momentum Modes}

Now we would like to discuss in some details the problem of the origin 
of the long lasting amplification of low-momentum pion modes seen 
in linear sigma model simulations, in particular in Ref. \cite{RW93}.

Let us first try to make a rough estimate of $\frac{\chi_0}{v}$ 
in their simulation. It is shown in Fig.2 of \cite{RW93} that 
$\langle \phi^2  \rangle$ is given by a constant superposed by 
an oscillatory component during the time scale of about 10-40 
in units of $a$ when large amplification takes place. 
Here $a$ is the lattice constant and it is taken as 
$a=(200\mbox{MeV})^{-1}$ in \cite{RW93}.
Since the frequency of the oscillatory component reflect the 
pion mass not the sigma mass it cannot be used to estimate 
the amplitude of oscillation in sigma direction. We thus interpret
(a quarter of) the constant component as a time and spatially averaged 
$\sigma$ oscillation. Since the low momentum modes are dominant at 
later times it may be utilized to estimate the amplitude of uniform 
background $\sigma$ oscillations in our scheme.
Using the constant component of $\sim 0.01 a^{-1}$ 
in $\langle \phi^2  \rangle$ we obtain $\frac{\chi_0}{v} \simeq 0.16$. 
It is a small number and indicates that our approximation is 
not totally absurd. But it is larger than 
$(\frac{m_{\pi}}{m_{\sigma}})^2 \simeq 0.05$, 
which implies that a new type of instability mentioned in Sec. III 
must also be excited.

Since we confine ourselves into 
$(\frac{\chi_0}{v}) < (\frac{m_\pi}{m_\sigma})^2$
in this paper, it is not possible to directly compare our parametric 
resonance mechanism to the Rajagopal-Wilczek simulation. 
Nevertheless, we attempt to make a bold comparison between these two 
by estimating the power spectrum, the corresponding 
quantity to the power calculated by them.  

Rajagopal-Wilczek used the definition of power spectrum \cite {Raja95}
\beq
P_{RW}(\vecvar{k},t)={1\over N^3}\left|\sum_{n_1,n_2,n_3}
e^{i\vecvar{k}\vecvar{n}a}
\phi(\vecvar{n},t)\right|^2,\label{powerRW}
\eeq
where $N$ indicates (one-dimensional) box size and they used $N=64$
and $\phi(\vecvar{n},t)$ denotes a single component of pion or $\sigma$ 
fields at lattice cite $\vecvar{n}$.
It has the dimension of $(mass)^2$ and hence the ordinates in their Fig.1 
must be understood to be plotted in units of $a^{-2}$.

We define the corresponding quantity, the power spectrum, in our 
framework. The natural definition is:
\beq
P_{ours}(\vecvar{k},t)=\langle |Q_{\vecvar{k}}(t)|^2\rangle.
\label{powerHM}
\eeq
The expectation value $\langle\cdot\rangle$ is to be 
evaluated by using the normalized squeezed state (\ref{squeezed}).
When the fields are confined into a finite volume $V$, 
Fourier integral in (\ref{defQ_k}) is modified to the discrete 
summation of the form
\beq
\xi(\vecvar{x},t)={1\over\sqrt{V}}\sum_ke^{-i\vecvar{k}\vecvar{x}}
Q_{\vecvar{k}}(t), 
\eeq
or inversely, 
\beq
Q_{\vecvar{k}}(t)={1\over\sqrt{V}}\int d^3xe^{i\vecvar{k}\vecvar{x}}
\xi(\vecvar{x},t).
\eeq

Then, using $V=(Na)^3$, our power spectrum  (\ref{powerHM}) is related 
with theirs via the following relation:
\beq
P_{RW}(\vecvar{k},t)=P_{ours}(\vecvar{k},t)/a^3.
\label{power_relation}
\eeq
It should be noticed that the number presented in \cite{RW93} is 
$P_{RW}(\vecvar{k},t) a^2$. Therefore, we must plot 
$P_{ours}(\vecvar{k},t)/a$ to compare our power with theirs.

We present our power spectra of pions and sigmas in Fig.7-10. 
We have adjusted some 
parameters so that they accord with the values adopted in \cite{RW93}; 
$v_0=87.4$MeV, $v=92.5$MeV, $m_\pi=135$MeV, though its effect is tiny 
and does not affect our conclusions. We employ the
amplitude parameter of background oscillation as $\chi_0/v=0.05$ in 
order to avoid the instability. 

In Fig.7 plotted is the power spectrum of pions at $t=0$.
One can see that there is a lot of power at relatively low momentum modes.
One may feel curious of the fact that the power is not small at $t=0$
at which we set the boundary condition of no particle production.
One can easily verify that it is natural because the latter condition 
implies that 
$Q \sim \frac{1}{\sqrt{2\Omega}} \sim \frac{1}{\sqrt{m_\pi}}$
and 
$\dot{Q} \sim \sqrt{\frac{\Omega}{2}} \sim \sqrt{m_\pi}$.

In Fig.8 we present the time evolution of power of pion field at $k=40$MeV, 
which is the longest wavelength case depicted in Fig.1 in \cite{RW93}
and it is where the largest amplification occurs. 
We observe that the average power is of order unity as we expect from 
the above estimation. It is a reasonable result because it is off resonance.

With these parameters, the first resonance appears in $k=268$MeV and 
the amplification of the power in this momentum bin is depicted in Fig.9. 
The $P_{ours}/a$ is growing by a factor of 50 and reaches to $\sim 20$ 
which is similar order of magnitude with Rajagopal-Wilczek's.
Of course, two numbers cannot be compared directly because of the 
difference involved between two computations including the initial 
conditions. Nevertheless, the qualitative feature 
of the growing power to the same order of magnitude may be an indication 
that the parametric resonance mechanism plays an important role in the 
long lasting amplification of low momentum modes.
One may argue that our results and Rajagopal-Wilczek's are qualitatively 
different because the amplification occurs in low momentum modes in their 
simulation, whereas it happens at resonance in our case. 
However, if we have not ignored the nonlinearity it might have mediated 
the enhancement at resonance to other modes. 

In Fig.10 plotted is the time evolution of power of sigma field at 
$k=40$MeV as in Fig.8. The power of sigmas is an order of magnitude 
smaller than that of pions, in agreement with the feature obtained in 
Rajagopal-Wilczek's simulation.

We now summarize our present understanding on possible origin of 
the long lasting amplification of low-momentum pion modes seen
in the linear sigma model simulations. We have suggested 
that the parametric resonance might be the cause.
The semi quantitative features of the power spectrum including 
its order of magnitude and the rate of growth is not inconsistent with 
our proposal.
We also pointed out that a new type of instability exists in 
our approximation scheme which, if real, would enhance the 
amplification of low-momentum modes. We, however, failed to achieve 
an intuitive understanding of the physical origin of the new instability.

\section{Conclusion and Discussion}

We have discussed the pion production via the parametric resonance 
mechanism within the linear sigma model. In particular, stimulated 
by the feature of the numerical simulations of the linear sigma model, 
we focused on the scenario in which classical background oscillations 
of the sigma model fields are in the $\sigma$ direction. Assuming 
small amplitude oscillation which may be natural in the late stage 
of evolution of DCC, we have shown that one can ignore effects due 
to nonlinearity and quantum back reactions for sufficiently long time. 
Thanks to this fact, we were able to construct an explicit quantum 
pion (and sigma) state which allows us to calculate the two particle 
correlations as well as the single particle distributions. 

Our treatment, though under very restrictive assumptions, may be good 
enough to illuminate characteristic features of the parametric resonance 
mechanism. We formulated the quantum theory of the system with 
Mathieu instability on a more conventional basis of quantum field theory 
and clarified the relationship between our and the formalism given by 
Shtanov, Traschen and Brandenberger \cite{STB95}.

We have analyzed the computed single particle distributions and 
clarified the structure of narrow resonances characteristic to the 
parametric resonance mechanism.
We then discussed the two pion correlations as a possible experimental 
probe for disoriented chiral condensate. Since the two pion correlations 
have unique characteristics, the back-to-back (in momentum space) 
correlations, it must give a clear signature which 
should merit the experimental hunting of DCC. In particular, it cannot 
be masked by the identical particle interference, the Hanbury-Brown-Twiss 
effect \cite{HBT}.

Why two particle correlation? It is certainly far more difficult to 
measure compared with the multiplicity distributions, on which all the 
recent experimental search for DCC rely \cite{minimax,WA98}. 
The global analysis using multiplicity distributions is powerful if 
large fractions of events are accompanied by the DCC domain formation. 
On the other hand, a different strategy is required for hunting if DCC 
is a rare phenomenon. 

We have discussed possible origin of 
the long lasting amplification of low-momentum pion modes
in the last section.
Our discussion cannot be a complete one, but we hope that it stimulates 
the readers' interests in this problem.

We should mention about the limitation and the drawback in our treatment 
in this paper. We ignored the quantum back reaction and the instability 
of $\sigma$ meson which should exist in the real world. It is the 
perfectly legitimate approximation if the amplitudes of background 
$\sigma$ oscillations are really small. 
However, it is possible that the amplitudes are sometimes large 
because of the prevailing thermal fluctuations in the initial stage. 
Since we are dealing with the system in which the coupling is 
really strong, most probably, the peak in the single particle 
distributions will go away after the quantum back reaction is taken 
into account.

Then, the key question is that if anything in our results remain valid 
after the quantum back reaction is taken into account. We argue that 
the answer is yes; it is the characteristic features of the two pion 
correlations which are discussed in detail in Sec. V.  
We now engage in a computation to verify our expectation.

How about the $\sigma\rightarrow\pi\pi$ coupling? It is clear that 
it also tends to smear out the resonance peaks and, more importantly, 
may obscure the signature of the back-to-back correlations. 
Again we need more elaborate treatment which includes the instability 
of $\sigma$ to make definitive statement about how much the signature 
survive in the case with $\sigma\rightarrow\pi\pi$ coupling.
The formalisms which may be suitable for such analysis have been 
investigated in detail \cite{boyan3,fred}.

There are also some recent proposals \cite {ACG98,Andreev}
that the hadronic medium effects may induce the similar back-to-back 
correlations in momentum space, which would mimic the signature 
of DCC discussed in this paper.
One would hope that it should be possible to find observational 
features which discriminate these two mechanisms. The task is, 
however, beyond the scope of this paper. 

\acknowledgements

During the long term investigation we benefited from numerous 
conversations in various stages of this work with 
Masayuki Asakawa, Dan Boyanovsky, Berndt M\"uller, Krishna Rajagopal, 
and Valery Rubakov.
In particular we thank Krishna for his informative and kind correspondences 
in comparing power spectra obtained by our and their calculations. 

The research of HM is partly supported by 
the Grant-in-Aid for International Scientific Research No. 09045036, 
Inter-University Cooperative Research, and 
the Grant-in-Aid for Scientific Research in Priority Areas No. 11127213, 
Ministry of Education, Science, Sports and Culture of Japan.

\section*{Appendix A}
\renewcommand{\theequation}{A.\themycounta}

We discuss in this Appendix some aspects of multiparticle correlations 
among particles involved in a normalized single-mode squeezed state 
\beqa
|\psi\rangle = \frac{1}
{\sqrt{\langle0|e^{\gamma^{*}K_{-}}e^{\gamma K_{+}}|0\rangle}}
\; e^{\gamma K_{+}}|0\rangle,
\eeq
where $K_{+}=\displaystyle\frac{1}{2}(a^{\dagger})^2$ and 
$K_{-}=\displaystyle\frac{1}{2}(a)^2$, and $\gamma$ denotes a 
complex number. 
The operators $K_{+}$ and $K_{-}$, together with $N=a^{\dagger}a$, 
form a closed algebra, 

\beqa
\left[K_{-}, K_{+}\right]  =  N + \frac{1}{2}, \qquad
\left[N, K_{\pm}\right]  =  \pm 2K_{\pm}, 
\eeq
which will play a role in our following computation. 

For convenience in the systematic treatment, we employ the generating 
function formalism developed by Koba, Nielson, and Olesen, \cite{KNO} 
and by Koba \cite{Koba}. 
The generating function is defined by 
\beqa
F[h] = \sum_{n=0}^{\infty} (1+h)^n P_n,
\eeq
using the multiplicity distribution $P_n=|\langle n|\psi\rangle|^2$ 
where
\beqa
|n\rangle = \frac{(a^\dagger)^n}{\sqrt{n!}}\;| 0 \rangle.
\eeq
The generating function can generate whole set of various moments when 
expanded in various different manner: 
\beqa
\begin{array}{ll}
F[h] &= \displaystyle{\sum_{n=0}^{\infty}\frac{h^k}{k!}F^{(k)}} \\
\\
     &= \displaystyle{\exp \left[\sum_{k=0}^{\infty}\frac{h^k}{k!}C^{(k)}
 \right]}.
\end{array}
\eeq
In these equations, $F(k)$ denotes 
\beqa
F^{(k)} = \langle n(n-1)(n-2)\cdots (n-k+1) \rangle,
\eeq
where $\langle \cdots \rangle$ implies the average over by the 
multiplicity distribution $P_n$; 
$\langle O \rangle \equiv \sum_{n=0}^{\infty}O_n P_n$. 
As is familiar in cluster expansion in statistical mechanics, $C^{(k)}$ 
represents the correlations 

\beqa
\begin{array}{ll}
C^{(1)} & =  \langle n \rangle, \\
&\\
C^{(2)} & =  \langle n(n-1) \rangle - \langle n \rangle^2, \\
&\\
C^{(3)} & =  \langle n(n-1)(n-2) \rangle -3 \langle n(n-1) \rangle 
\langle n \rangle - \langle n \rangle^3,
\end{array}
\eeq
and so on. 

The expression of the generating function in a form of operator expectation 
value is give by Koba \cite{Koba}; 
\beqa
F[h] = \langle \psi | : e^{h a^{\dagger} a}: |\psi \rangle,
\eeq
where $::$ indicates to take the normal ordering. It proves to be a 
very useful formula for our purpose. 
Toward calculating $F[h]$, we define the quantity $A_m$ as 
$A_m \equiv \langle \psi | (a^\dagger)^m (a)^m | \psi \rangle$. 
Using the algebra applied to the vacuum state 
\beqa
\begin{array}{ll}
\left[ a^m, e^{\gamma K_{+}} \right]|0\rangle & =  
\gamma a^{m-1}a^\dagger e^{\gamma K_{+}} |0\rangle, \\
&\\
\langle 0| \left[e^{\gamma^{*}K_{-}}, (a^{+})^m  \right] & =
\gamma^* \langle 0 | e^{\gamma^{*}K_{-}} a(a^\dagger)^{m-1},
\end{array}
\eeq
one can derive the recursion relation among $A_m$; 
\beqa
\begin{array}{ll}
A_0 &= 1,\\
&\\
A_1 &= \langle n \rangle, \\ 
&\\
A_{m+2} &= \langle n \rangle \biggl[(2m+3)A_{m+1} + (m+1)^2 A_m \biggr]
\quad (m\geq0).
\end{array}
\label{recursion}
\eeq
One can show, by direct computation, that
\beqa
\begin{array}{ll}
\langle n \rangle &= \langle \psi | a^\dagger a | \psi \rangle \\
&\\
&= \displaystyle{\frac{|\gamma|^2}{1-|\gamma|^2}}.
\end{array}
\eeq
We can convert the recursion relation (\ref{recursion}) into the 
differential equation obeyed by the generating function
\beqa
\left[(h^2+2h-\frac{1}{\langle n \rangle})\frac{d^2}{dh^2} + 
3(h+1)\frac{d}{dh} +1\right] F[h] =0.
\eeq
We make a change of variable 
\beqa
t= \frac{\langle n \rangle}{1+ \langle n \rangle}(1+h)^2,
\eeq
to transform the differential equation into the standard form 
\beqa
\left[ t(1-t)\frac{d^2}{dt^2} + (\frac{1}{2} -2t)\frac{d}{dt}
- \frac{1}{4} \right] F[t] = 0.
\eeq
The solution to the equation subject to the boundary conditions 

\beqa
\begin{array}{ll}
\displaystyle
F[t]\mid_{t=\frac{\langle n \rangle}{1+\langle n \rangle}} 
&= 1, \\
&\\
F'[t]\mid_{t=\frac{\langle n \rangle}{1+\langle n \rangle}} 
&= \displaystyle\frac{1+\langle n \rangle}{2}, \\
\end{array}
\eeq
is uniquely given by 
\beqa
F[t]= {1\over\sqrt{(1+\langle n \rangle )(1 - t)}}.
\eeq
The boundary conditions correspond to $A_0 = 1$, $A_1 = \langle n \rangle$, 
and $A_2 = \langle n \rangle(3\langle n \rangle + 1)$.  
Thus, we have obtained the generating function as 
\beqa
F[h] = \frac{1}{\sqrt{1-\langle n \rangle (h^2+2h)}}.\label{gene}
\eeq
The correlation moments can be easily computed as follows: 

\beqa
\begin{array}{ll}
C^{(1)} & = \langle n \rangle, \\
&\\
C^{(2)} & = \langle n \rangle (2\langle n \rangle + 1),\\
&\\
C^{(3)} & = 2\langle n \rangle^2 (4\langle n \rangle + 1),\\
&\\
C^{(4)} & = 48\langle n \rangle^2 \left(\langle n \rangle^2 + \langle n 
\rangle + 
\displaystyle\frac{1}{8}\right).\\
\end{array}
\eeq

Since the generating function (\ref{gene}) is the function of $(1+h)^2$,
the multiplicity distribution $P_n$ for odd $n$ vanishes. That for 
even $n$ is given by 
\beqa
P_{2k} = \frac{(2k)!}{2^{2k}(k!)^2}\frac{1}{\sqrt{1+\langle n \rangle}}
\left(\frac{\langle n \rangle}{1+ \langle n \rangle}\right)^k,
\label{binomial}
\eeq
which is nothing but the negative binomial distribution familiar in 
hadronic multiparticle phenomenology. The expression (\ref{binomial}) 
reproduces the results obtained by Yoshimura \cite{yoshimura} 
who also calculated all 
the off-diagonal elements of the density matrix. 

For completeness, we calculate the KNO scaling function for the multiplicity 
distribution (\ref{binomial}). 
It is obvious, from the expression of the correlation moments, that 
the squeezed state has ``long range'' correlations 
\beqa
\frac{C^{k}}{\langle n \rangle^k} 
\mathop{\longrightarrow}_{\langle n \rangle \rightarrow \infty} \mbox{finite},
\eeq
the multiplicity distribution obeys the KNO scaling behavior. 
Either by solving the moment problem or by using the explicit form 
(\ref{binomial}), one can show that 
\beqa
\begin{array}{ll}
\psi(z) &= \lim_{\langle n \rangle \rightarrow \infty} 
\langle n \rangle P_n \\
&\\
&= \displaystyle\sqrt{\frac{2}{\pi z}}e^{-z/2} 
\end{array}
\eeq

\section*{Appendix B}
\renewcommand{\theequation}{B.\themycountb}

We derive relations among two-pion correlation functions of various 
different charge states which follow from isospin invariance. 
We restrict ourselves into the case with iso-singlet state which 
is of concern to us. 
We also confine ourselves into the case of zero momentum pions; 
the treatment for nonzero momentum pions is different and is much 
more involved. 

We work with the (slightly modified) Horn-Silver \cite{HS}
representation of the isospin generator
\beqb
\vec{T} = {a^i}^{\dagger} \vec\tau_{ij}{a^j},\hspace{1cm} (i,j = +,-,0)
\eeq
where
\beqb
\tau_1 = \displaystyle\frac{1}{\sqrt 2}
\left[
\begin{array}{ccc}
0 & 0 & 1\\
0 & 0 & 1\\
1 & 1 & 0
\end{array}
\right],
\hspace{1cm}
\tau_2 = \displaystyle\frac{1}{\sqrt 2}\left[
\begin{array}{ccc}
0 & 0 & -i \\
0 & 0 & i \\
i & -i & 0 
\end{array}
\right],
\hspace{1cm}
\tau_2 = \left[
\begin{array}{ccc}
1 & 0 & 0 \\
0 & -1 & 0 \\
0 & 0 & 0 
\end{array}
\right].
\eeq
The explicit form of $\vec{T}$ is as follows:
\beqb
\begin{array}{ll}
T^{1} &= \displaystyle\frac{1}{\sqrt 2}\left[
\ac{+}\aa{0}+\ac{0}(\aa{+}+\aa{-})+\ac{-}\aa{0} \right],\\
&\\
T^{2} &= \displaystyle\frac{i}{\sqrt 2}\left[
-\ac{+}\aa{0}+\ac{0}(\aa{+}-\aa{-})+\ac{-}\aa{0} \right],\\
&\\
T^{3} &= \displaystyle\frac{1}{\sqrt 2}\left[
\ac{+}\aa{+}-\ac{-}\aa{-}\right],
\end{array}
\eeq
where $\aa{+}, \aa{0},$ and $\aa{-}$ are the annihilation operators 
of $\pi^{+}, \pi^{0},$ and $\pi^{-}$, respectively, as defined in 
(\ref{chargestate}). The relation between the charge eigenstate 
operators and the adjoint representation operators are given in 
(\ref{chargestate}).

Let $|\psi \rangle$ be the iso-singlet state.
Then, it is annihilated by the isospin operator, 
$\vec{T}|\psi \rangle =0$. From this, it follows that 
\beqb
\langle \psi | e^{i\vec{\xi}\cdot\vec T} {a^k}^{\dagger}{a^l}| \psi\rangle = 
\langle \psi | {a^k}^{\dagger} a^l | \psi \rangle,
\eeq
for any $k,l$.
This is the relation that should be obeyed for all orders in $\vec{\xi}$. 
If we take the first-order term in $\vec{\xi}$, we obtain the sum rule
\beqb
\langle \psi |\vec T {a^k}^{\dagger} a^l | \psi\rangle = 0 \hspace{1cm}
(\mbox{for any } k,l).
\eeq
Due to isospin invariance, it is easy to show that 
$\langle \pi^{+} \rangle = \langle \pi^{-} \rangle = \langle \pi^{0}
\rangle$. Then, the $T^1$ (or $T^2$) sum rule produces the relations 
like  
\beqb
\langle \pi^{+}\pi^{0} \rangle = \langle \pi^{-}\pi^{0} \rangle,
\eeq
which also follows from the $T^3$ sum rule. 
The useful relations comes from $T^3$-sum rule 
\beqb
\langle \pi^{+}\pi^{+} \rangle = \langle \pi^{+}\pi^{-} \rangle = 
\langle \pi^{-}\pi^{-} \rangle.
\eeq
In these equations, $\langle \pi^{\alpha}\pi^{\beta} \rangle$ implies 
the two-pion distribution of charge state $\alpha$ and $\beta$. 

The two-pion correlation function may be defined as 
\beqb
\begin{array}{ll}
C(\pi^{+},\pi^{+}) &= 
\langle \pi^{+}\pi^{+} \rangle - \langle \pi^{+} \rangle 
- \langle \pi^{+} \rangle^2, \\
&\\
C(\pi^{+},\pi^{0}) &= 
\langle \pi^{+}\pi^{0} \rangle - \langle \pi^{+} \rangle 
\langle \pi^{0} \rangle,\\
&\\
\mbox{etc.} \hspace{1cm}&
\end{array}
\eeq
Then, we obtain the isospin sum rule, 
\beqb
C(\pi^{+}, \pi^{-}) = C(\pi^{+}, \pi^{+}) + \langle \pi \rangle,
\eeq
where $\langle \pi \rangle \equiv \langle \pi^{+} \rangle = 
\langle \pi^{-}\rangle = \langle \pi^{0}\rangle$. 
Since $\langle \pi\rangle$ is positive, the inequality among two-pion 
correlations follows; 
\beqb
C(\pi^{+}, \pi^{-}) > C(\pi^{+}, \pi^{+})
\eeq
It may be surprising that the unlike-sign correlation is stronger than 
like-sign correlation. But it is the consequence of the isospin invariance 
with iso-singlet nature of the multipion state.


\begin{figure}
\caption{The single pion momentum distribution as a function of the scaled 
time $z={1\over 2}(m_\sigma t+\pi)$. The amplitude of oscillation is taken
as $\chi_0/v=0.025$.}\label{amp0.025}
%
\caption{The single pion momentum distribution as a function of the scaled 
time $z={1\over 2}(m_\sigma t+\pi)$. The amplitude of oscillation is taken
as $\chi_0/v=0.05$.}\label{amp0.05}
%
\caption{The single pion momentum distribution as a function of the scaled 
time $z={1\over 2}(m_\sigma t+\pi)$. The amplitude of oscillation is taken
as $\chi_0/v=0.1$.}
%
\caption{The single sigma momentum distribution as a function of the scaled 
time $z={1\over 2}(m_\sigma t+\pi)$. The amplitude of oscillation is taken
as $\chi_0/v=0.05$.}\label{s-amp0.05}
%
\caption{The pion identical momentum correlation function is depicted.
The amplitude of oscillation is taken as $\chi_0/v=0.05$.}\label{csame}
%
\caption{The pion back-to-back momentum correlation function is depicted.
The amplitude of oscillation is taken as $\chi_0/v=0.05$.}\label{copp}
%
\caption{The power spectrum of pion at $t=0$ is depicted. Due to the 
boundary condition of no particle production at $t=0$, the power spectrum
is not small.}
%
\caption{The time evolution of power spectrum of pion field at 
$k=40$MeV is presented. It corresponds to the longest wavelength bin 
in Fig. 1 of Rajagopal and Wilczek's paper.}
%
\caption{The time evolution of power spectrum of pion field at 
momentum $k=268$MeV, which corresponds to the first resonance 
band, is drawn.}
%
\caption{The time evolution of power spectrum of sigma field at 
$k=40$MeV is depicted.}
\end{figure}
%
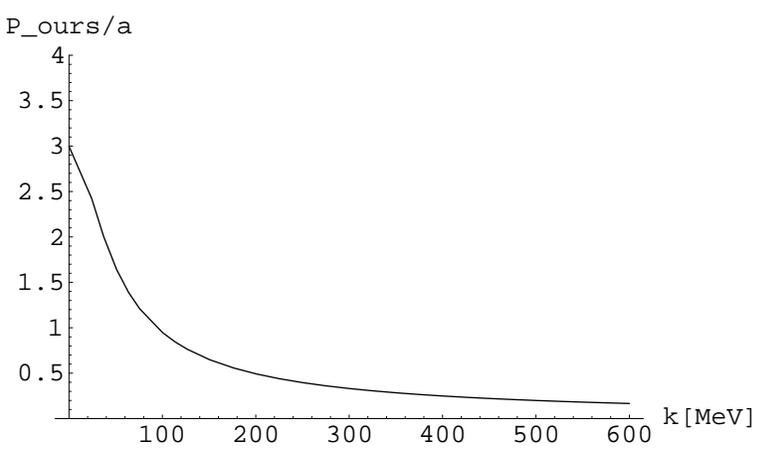
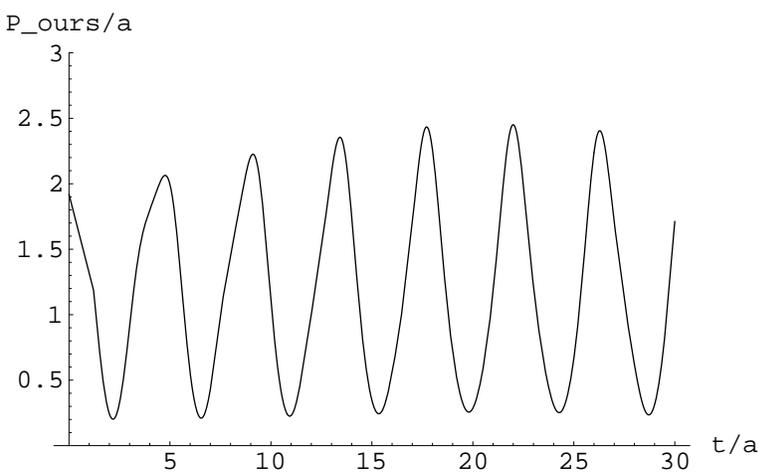
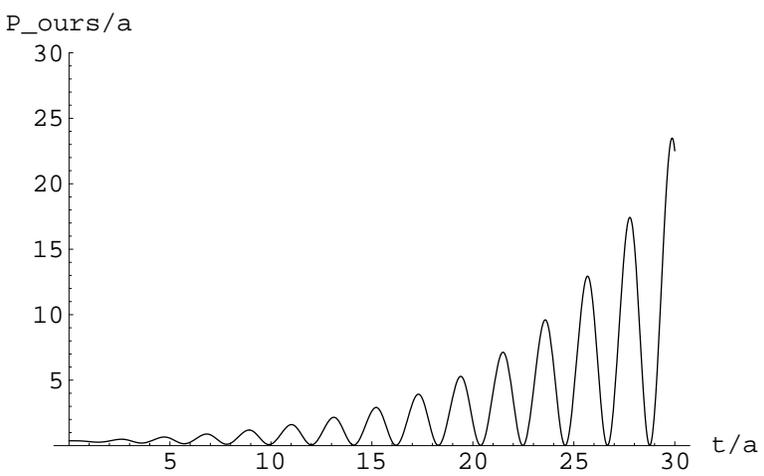
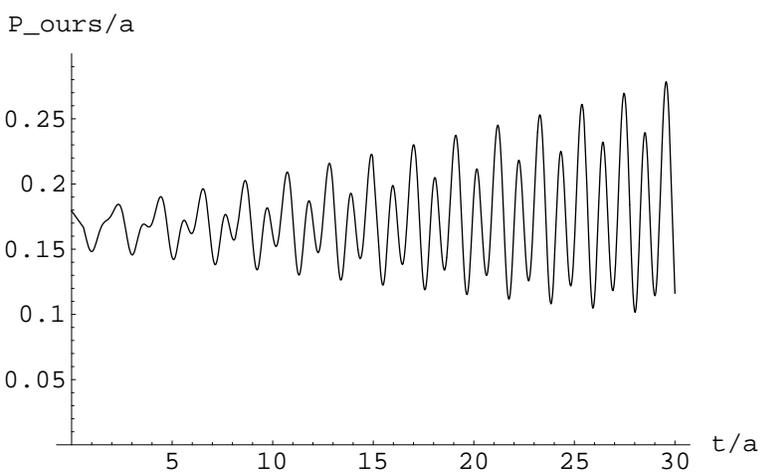
\end{document}